\documentclass[notoc]{JHEP3}
\usepackage{amsmath,amssymb}
\usepackage{graphicx}

\newcommand{\be}{\begin{equation}}
\newcommand{\ee}{\end{equation}}
\newcommand{\bea}{\begin{eqnarray}}
\newcommand{\eea}{\end{eqnarray}}
\newcommand{\beann}{\begin{eqnarray*}}
\newcommand{\eeann}{\end{eqnarray*}}

\newcommand{\ba}{\begin{array}}
\newcommand{\ea}{\end{array}}

\newcommand{\p}{\partial}

\title{ On non-stationary Lam\'e equation from 
WZW model and spin-1/2 XYZ chain
}

\author{
Ta-Sheng Tai$^{1,2}$ and Reiji Yoshioka$^2$\\

$^1$Interdisciplinary Graduate School of Science and Engineering\\
Kinki University\\
Osaka 577-8502, Japan\\

$^2$Advanced Mathematical Institute\\
Osaka City University\\ 
Osaka 558-8585, Japan\\

\\
\\
{\tt tasheng} {\rm at} {\tt alice.math.kindai.ac.jp},
{\tt yoshioka} {\rm at} {\tt sci.osaka-cu.ac.jp}
}

\abstract{
We study the link between WZW model and the 
spin-1/2 XYZ chain. 
This is achieved by comparing the second-order differential equations from them. 
In the former case, the equation is 
the Ward-Takahashi identity satisfied by one-point toric conformal blocks.
In the latter case, it arises from Baxter's $TQ$ relation. 
We find that the dimension of the representation space w.r.t. 
the $V$-valued primary field in these conformal blocks gets 
mapped to the total number of chain sites. 
By doing so, Stroganov's 
``The Importance of being Odd" 
(cond-mat/0012035) can be consistently understood in terms of WZW model language. 
We first confirm this correspondence 
by taking 
a trigonometric limit of the XYZ chain. 
That eigenstates of the resultant two-body 
Sutherland model from Baxter's $TQ$ relation 
can be obtained by deforming toric conformal blocks 
supports our proposal. }

\begin{document}


\section{Introduction}\label{sec1}

About twenty years ago, a series of pioneering papers \cite{Ba1, BF, FFR} established 
an intriguing connection between XXX Gaudin and Wess-Zumino-Witten (WZW) models. 
That is, the problem of diagonalizing 
commuting Hamiltonians%
\footnote{Their simultaneous diagonalization is 
solved by algebraic Bethe ansatz \cite{KBI} 
 and Sklyanin's separation of variables \cite{SK}.
Two approaches are essentially equivalent and amount to considering 
the quantized Gaudin spectral curve.}
 of XXX Gaudin model is 
translated into solving Knizhnik-Zamolodchikov (KZ) equations defined on ${\bf CP}^1$ \cite{KZ}. 
Indeed, Bethe roots of Bethe ansatz equations in the 
inhomogeneous XXX Gaudin model 
turn out to constitute solutions to 
KZ equations at critical level.
 Later on, the authors of 
\cite{KT1, BPL, KT2} further extended this direction to the elliptic case. 
Certainly, their works are based on important investigations on both XYZ Gaudin model \cite{TS} and conformal field theory (CFT) on elliptic curves \cite{B,EK, Suzuki:1994hv}. 

\begin{figure}[tt]
\begin{center}
\includegraphics[width=0.8\textwidth]
{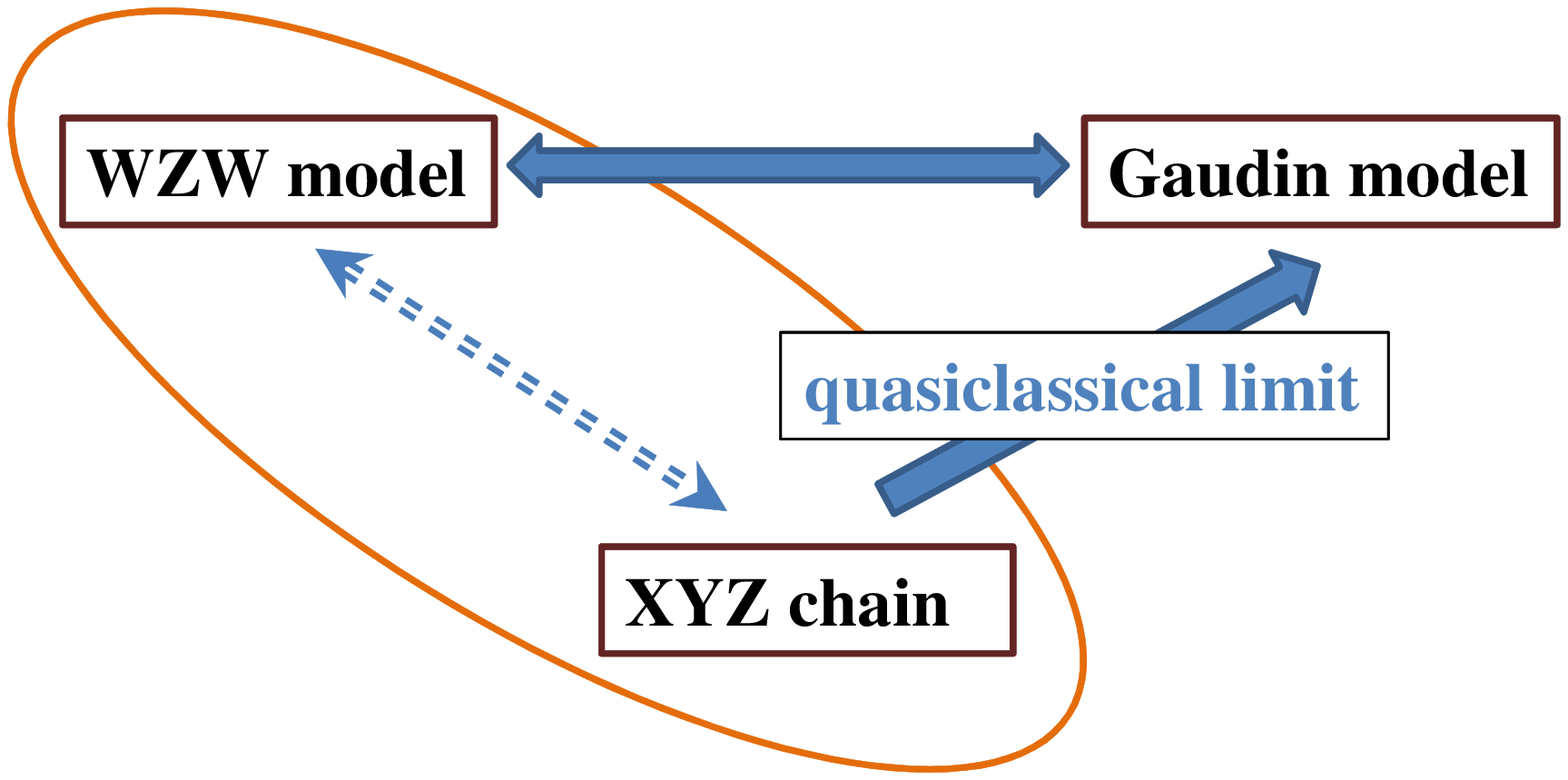}
\end{center}
\caption{Two solid arrows above represent the known connections between spin-chain
and WZW models. 
The encircled part indicates the novel relationship under consideration. 
\label{F1}
}
\end{figure}

In this letter, we would like to add into the above picture a novel 
element: 
a relation between WZW model and the spin-1/2 XYZ chain as depicted 
in Fig. \ref{F1}. 
By examining non-stationary Lam\'e equations on both sides we are able to interpret Stroganov's proposal 
(The Importance of being Odd) 
\cite{TT-St} 
from the viewpoint of CFT under the dictionary listed in Table \ref{dic}. 
\begin{table}[htb]
\label{dic}
\begin{center}
\caption{Dictionary}
\begin{tabular}{lll}
\hline
& {\bf Spin-1/2~XYZ~chain} & {\bf WZW~model}\\
\hline\hline
{\bf non-stationary Lam\'e eq.}& Baxter's $TQ$ eq.
&KZB eq. (WT identity)\\
\hline
 {\bf coupling~const.} & site~number & dim.~of~
$\mathfrak{sl}_2$~rep.  \\
\hline
{\bf time} & anisotropy parameter&torus~moduli\\
\hline
{\bf space} & spectral~parameter&Cartan~moduli\\
\hline
\end{tabular}
\end{center}
\end{table}

More precisely, in \cite{TT-RS} 
Razumov and Stroganov made a conjecture about the exact ground-state 
eigenvalue of the transfer matrix in the spin-1/2 XYZ chain. This conjecture 
holds only for the odd chain site number and plays a crucial role in 
deriving the aforementioned Lam\'{e} equation \cite{BM}. 
On the other hand, one-point toric conformal blocks 
exist only when the 
dimension of the $\mathfrak{sl}_2$ representation 
space w.r.t. the inserted primary field is odd. 
It is thus tempting to connect these two facts through Table \ref{dic}. 
As a test, we perform a trigonometric degeneration of the XYZ chain.
Consequently, that eigenstates of 
Sutherland-type equations descending from Baxter's $TQ$ relation 
reduce to Schur polynomials under certain limit 
is well reflected by imposing a corresponding constraint on 
WZW toric conformal blocks.

We organize this letter as follows.
In the next section, we review how Lam\'e equations emerge 
from the spin-1/2 XYZ chain as a result of 
Baxter's $TQ$ relation. We compare it with 
Knizhnik-Zamolodchikov-Bernard (KZB) equations in section 3. 
In section 4, we justify this comparison via a 
trigonometrical reduction. Finally, 
a summary is given in section 5.

\section{Spin-1/2 XYZ chain side}\label{sec2}

Let us briefly review how the non-stationary Lam\'{e} equation is
obtained from Baxter's $TQ$ equation of the
spin-1/2 XYZ chain \cite{TT-RS, BM} whose Hamiltonian is 
described by 
\begin{align}
H_{XYZ}= \sum_{n=1}^M \biggl\{
J_X S_n^X S_{n+1}^X+ J_Y S_n^Y S_{n+1}^Y+ J_Z S_n^Z S_{n+1}^Z \biggl\}.
\label{TT-HAM}
\end{align}
Here, $S_n^{X,Y,Z}=\sigma_n^{X,Y,Z}/2$
($\sigma_n^{X,Y,Z}$: Pauli matrix) acts on the $n$-th site and 
the periodic boundary condition $S_{M+1}^{X,Y,Z}=S_1^{X,Y,Z}$ is imposed. 
Recall that the 
terminology XYZ means anisotropic $J$'s 
while the partial anisotropy $J_X=J_Y \ne J_Z$
(isotropy $J_X=J_Y = J_Z$) case is called the XXZ (XXX) chain. $H_{XYZ}$ acts on the tensor product
$V_1 \otimes V_2 \otimes \cdots \otimes V_M$
where each $V_n$ is a complex
two-dimensional space ${\bf C}^2$ spanned by the up- and down-spin states.

A fundamental ingredient in integrable spin-chain models is the
$R$ matrix. For the spin-1/2 XYZ chain,
its matrix elements are given by
\[
R(z)
=
\begin{pmatrix}
a(z)  & 0 & 0 & d(z) \\
0 & b(z) & c(z) & 0 \\
0 & c(z) & b(z) & 0 \\
d(z) & 0 & 0 & a(z)
\end{pmatrix}
\]
where (nome: $q=e^{ \pi i \tau}$)
\begin{align}
a(z)&=\rho \theta_4(2\eta|q) \theta_4(z|q) \theta_1(z+2\eta|q), \ \ \ \
b(z)=\rho \theta_4(2\eta|q) \theta_1(z|q) \theta_4(z+2\eta|q), \nonumber \\
c(z)&=\rho \theta_1(2\eta|q) \theta_4(z|q) \theta_4(z+2\eta|q), \ \ \ \
d(z)=\rho \theta_1(2\eta|q) \theta_1(z|q) \theta_1(z+2\eta|q), \nonumber \\
\rho&=\frac{2}{\theta_2(0|q^{\frac{1}{2}}) \theta_4(0|q)},  \nonumber \\
\theta_1(z|q)&=-i \sum_{n\in {\bf Z} } (-1)^n q^{(n+\frac{1}{2})^2}\mathrm{e}^{(2n+1)\pi iz}, \ \ \ \ ~~~~~~~
\theta_2(z|q)= \sum_{n\in {\bf Z} }  q^{(n+\frac{1}{2})^2}
\mathrm{e}^{(2n+1)\pi iz}, \nonumber \\
\theta_3(z|q)&= \sum_{n\in {\bf Z} }  q^{n^2}\mathrm{e}^{2n\pi iz}, \ \ \ \ ~~~~~~~~~~~~~~~~~~~~~~~~~~~~
\theta_4(z|q)= \sum_{n\in {\bf Z} } (-1)^n q^{n^2}
\mathrm{e}^{2n\pi iz}. \nonumber
\end{align}
Note that $(q, \eta)$ determines the anisotropy
parameters of the XYZ chain through Jacobi's elliptic functions:
\begin{align}
J_X=1+{\bf k} \mathrm{sn}^2 (\pi \eta,{\bf k}),  \ \ \ J_Y=1-
{\bf k} \mathrm{sn}^2 (\pi \eta,{\bf k}),  \ \ \
J_Z=\mathrm{cn} (\pi \eta,{\bf k}) \mathrm{dn} (\pi \eta,{\bf k}),  \ \ \
{\bf k}=\frac{\theta^2_2(0|q)}{\theta^2_3(0|q)}.
\nonumber
\end{align}
Also, $z$ denotes the spectral parameter which plays an important role
in quantum integrable models.
When $q \to 0$, due to ${\bf k} \to 0$ as well as 
\begin{align}
   \mathrm{sn} (\pi \eta,{0})=\sin \pi \eta, \ \ \ \ \ \
\mathrm{cn} (\pi \eta,{0})=\cos \pi \eta,  
 \ \ \ \ \ \  \mathrm{dn} (\pi \eta,{0})=1,
\nonumber
\end{align}
one yields a XXZ chain with  
\begin{align}
J_X=J_Y=1, \ \ \ \ \ \ J_Z=\cos \pi \eta.
\nonumber
\end{align}
Remark that $2J_Z={\bf q}+{\bf q}^{-1}$ where ${\bf q}=\exp(\pi i \eta)$ is 
referred to as the deformation parameter ${\bf q}$ of the quantum group $U_{\bf q} ({\mathfrak{sl}}_2)$.

In fact, three $R$-matrices acting on $V_1 \otimes V_2 \otimes V_3$
satisfy the famous Yang-Baxter relation:
\begin{align}
R_{12}(z)R_{13}(z+w)R_{23}(w)
=R_{23}(w)R_{13}(z+w)R_{12}(z). \nonumber
\end{align}
The subscript of, say, $R_{13}(z)$ 
means that it acts on $V_1 \otimes V_3$. 
From these $R$-matrices, one can construct the monodromy matrix $T_a(z)$
acting on $V_a \otimes ( V_1 \otimes \cdots \otimes V_M)$:
\begin{align}
\nonumber
T_{a}(z)=R_{aM}(z) \cdots R_{a1}(z)
=\left(
\begin{array}{cc}
A(z)  & B(z) \\
C(z) & D(z)
\end{array}
\right).
\end{align}
One can further yield
the transfer matrix ${\mathcal T}(z)=\mathrm{tr}_a T_a(z)$ by
performing a trace over the auxiliary space $V_a$.
Utilizing the above Yang-Baxter relation repeatedly,
one arrives at the so-called $RTT$ relation:
\begin{align}
\nonumber
R_{ab}(z-w)T_a(z)T_b(w)
=T_b(w)T_a(z)R_{ab}(z-w),
\nonumber
\end{align}
from which the commutativity of transfer matrices follows:
\begin{align}
[ \mathcal{T}(z), \mathcal{T}(w) ]=0. \nonumber
\end{align}
Let us briefly explain why 
there exists a common ${\bf q}$ between the XXZ Hamiltonians $H_{XXZ}$ and $U_{\bf q} ({\mathfrak{sl}}_2)$ encountered above. 
First, one can construct the XXZ transfer matrix from a product of 
$R$-matrices of the affine quantum group $U_{\bf q} (\widehat{\mathfrak{sl}}_2)$. Then, in order 
to derive $H_{XXZ}$ the standard way is to take the logarithmic derivative of the 
XXZ transfer matrix.

\subsection{Baxter's $TQ$ relation as non-stationary Lam\'{e} equation}
Baxter's $Q$-operator method
is a powerful tool for finding the eigenvalue of transfer matrices.
Let us briefly sketch his approach here.
One prepares a local matrix $S_{aj}(z)$ which acts on
$W_a \otimes V_j$ where $W_a=\mathbf{C}^L$ when $\exp(\pi i\eta L)=1$. 
From $S_{aj}(z)$ we construct a global matrix
\begin{align}
Q_{a}(z)=&S_{aM}(z) \cdots S_{a1}(z) \nonumber
\end{align}
acting on
$W_a \otimes ( V_1 \otimes \cdots \otimes V_M)$.
Baxter's $Q$-operator is defined by
$\mathcal{Q}(z)=\mathrm{tr}_{W_a} Q_a(z)$ which acts also on the previous
$V=V_1 \otimes \cdots \otimes V_M$.
Baxter's idea was to consider the product of $\mathcal{T}(z)$ and
$\mathcal{Q}(z)$
\begin{align*}
\mathcal{T}(z)\mathcal{Q}(z)&=\mathrm{tr}_{V_a \otimes W_{a^\prime}}
\Bigg\{
\prod_{j=1}^M R_{aj}(z)S_{a^\prime j}(z)
\Bigg\} \\
&=
\mathrm{tr}_{V_a \otimes W_{a^\prime}}
\Bigg\{\prod_{j=1}^M
U R_{aj}(z)S_{a^\prime j}(z) U^{-1}
\Bigg\}
\end{align*}
followed by a gauge transformation: $R_{aj}(z)S_{a^\prime j}(z) \to
U R_{aj}(z)S_{a^\prime j}(z) U^{-1}$ such that
the latter becomes a triangular matrix via a suitable $U$.
By doing so, both eigenvalues of
$\mathcal{T}(z)$ and $\mathcal{Q}(z)$
are shown to satisfy Baxter's $TQ$ relation \cite{TT-Baxter1}%
\footnote{See \cite{11,12,13,14,15} for recent applications of
Baxter's $TQ$ relation
to 4d gauge theories on $\Omega$-backgrounds and Nekrasov's partition
function \cite{16}.}
\begin{align}
{\mathcal T}(z)\mathcal{Q}(z)=\phi(z-\frac{\eta}{2})\mathcal{Q}(z+ \eta)
+\phi(z+\frac{\eta}{2}) \mathcal{Q}(z- \eta)
\label{TT-TQ}
\end{align}
with $\phi(z)=\theta_1^M(z|q)$.

At the Razumov-Stroganov point%
\footnote{That $\eta$ differs from the typical value $\pi/3$ is 
 due to our choice of 
two half-periods $(\omega_1, \omega_2)=(1/2,  \tau/2)$ of 
Weierstrass's elliptic function 
$\wp(z|q)\equiv \wp(z|\omega_1, \omega_2)$ instead of $(\pi/2, \pi \tau/2)$. These two notations are related by 
$\wp(tz| t\omega_1, t\omega_2)=t^{-2} \wp(z|\omega_1, \omega_2)$.} 
 $\eta=1/3$ \cite{TT-RS}, 
a particularly simple expression for the 
ground-state eigenvalue of $\mathcal{T}(z)$ 
was conjectured to be $\phi(z)$ \cite{TT-St,TT-RS, TT-Baxter2}.
Their conjecture holds only
when the number of chain sites is odd: $M=2n+1$ ($n \in \mathbf{Z}_{\ge 0}$).
Inserting this ${\mathcal T}(z)$ into \eqref{TT-TQ},
Bazhanov and
Mangazeev \cite{BM} managed to show that $Q$-operators dressed by
\begin{align}
\nonumber
\Psi_{\pm}^{(8 \mathrm{vertex})}(z,q,n)=
\frac{\theta_1^{2n+1}(z|q)}{\theta_1^n(3z|q^3)} \mathcal{Q}_{\pm}(z,q,n)
\end{align}
satisfy the non-stationary Lam\'{e} equation:
\begin{align}
 6 q \frac{\partial}{\partial q} \Psi_{\pm}^{(8 \mathrm{vertex})} (z,q,n)
 = \frac{1}{\pi^2}\left\{-\frac{\partial^2}{\partial z^2}
 + {9} n(n+1) \wp(3z|q^3) + c(q,n) \right\}
\Psi_{\pm}^{(8 \mathrm{vertex})}(z,q,n).
 \label{TT-Lame1}
\end{align}
Let two half-periods of Weierstrass's elliptic function
$\wp(z|q)\equiv \wp(z|\omega_1, \omega_2)$ 
be $(\omega_1, \omega_2)\equiv (1/2, \tau/2)$. Then,
\begin{align*}
\wp(z|q)&=-\zeta^\prime(z|q), \\
\zeta(z|q)&=\frac{\theta_1^\prime(z|q)}{\theta_1(z|q)}+2 \eta_1(q) z, \\
\eta_1(q)&=4\pi^2
\left(\frac{1}{24} -\sum_{n=1}^\infty \frac{n q^n}{ 1-q^n} \right)
=-\frac{1}{6} \frac{\theta_1^{\prime \prime \prime}(0|q)}
{\theta_1^{\prime}(0|q)}, \\
c(q,n)&=18n(n+1) \eta_1(q^3).
\end{align*}

In terms of the new variable $s=3z$,
one can rewrite \eqref{TT-Lame1} into
\begin{align}
 \frac{2}{3} q \frac{\partial}{\partial q} \Psi_{\pm}^{(8 \mathrm{vertex})}
(s/3,q,n)
 = \frac{1}{\pi^2}\left\{-\frac{\partial^2}{\partial s^2}
 + n(n+1)\Big( \wp(s|q^3) +2 \eta_1(q^3)\Big) \right\}
\Psi_{\pm}^{(8 \mathrm{vertex})}(s/3,q,n).
\label{TT-Lame2}
\end{align}
We can replace $ 2q {\partial}/3{\partial q}$ by 
$2{\partial}/ \pi i {\partial \bar{\tau}}$ with
$3\tau=\bar{\tau}$.

\section{WZW model side}\label{sec3}

Our goal is to see the appearance of \eqref{TT-Lame1} within the context of WZW model and 
then interpret Stroganov's claim geometrically.

\subsection{Affine Lie algebra}
The conformal symmetry here will be realized by means of the level-$k$ affine Lie algebra $\widehat{\mathfrak{g}}$. 
In general, 
the integrable irreducible $\widehat{\mathfrak{g}}$-module 
$L_{k, \lambda}$ is characterized by a set of non-negative 
highest weights 
${\lambda_a}$ $(a=0,\cdots,r={\rm rank})$ w.r.t. $\mathfrak{g}$ 
(simple finite-dimensional Lie algebra) where 
$ \lambda_0=k-(\theta,\lambda) \ge 0$. 
Symbolically, 
\begin{align*}
\lambda_a \in P^k_+=\{ \lambda_a \in P_+~|~0\le (\theta,\lambda)\le k
\},  ~~~~~~~~a=1,\cdots,r.
\end{align*}
$L_{k, \lambda}$ with all null states being decoupled 
forms an unitary representation of $\widehat{\mathfrak{g}}$.%
\footnote{Let $\widehat{\mathfrak{g}}=
\widehat{\mathfrak{n}}_-\oplus 
\widehat{\mathfrak{h}} \oplus \widehat{\mathfrak{n}}_+$. 
Given the 
highest weight state $|\lambda\rangle$ which is 
annihilated by generators in $\widehat{\mathfrak{n}}_+={\mathfrak{n}_+} \oplus 
({\mathfrak{g}} \otimes {\bf C}[z])$, 
the reducible module is gained by applying to $|\lambda\rangle$ repeatedly 
generators in 
$\widehat{\mathfrak{n}}_-={\mathfrak{n}_-} \oplus 
({\mathfrak{g}} \otimes {\bf C}[z^{-1}])$. In order to decouple null 
states from the module, one must further impose 
\begin{equation*}
(E_\theta   \otimes z^{-1} )^{\lambda_0 +1} |\lambda\rangle=0
,  ~~~~~~~~(E_{-\alpha})^{(\alpha^{\vee}, \lambda) +1} |\lambda\rangle=0
,  ~~~~~~~~\theta:~\rm{highest~root}.
\end{equation*}}
Let us proceed to explain various notations used above.

We focus only on the $A_{N-1}$-type Lie algebra $\mathfrak{sl}_N$ whose $N^2-1$ 
generators can get triangularly decomposed into 
$\mathfrak{sl}_N =\mathfrak{n}_-\oplus \mathfrak{h} \oplus \mathfrak{n}_+$ ($\mathfrak{h}$: Cartan 
subalgebra, $r={\rm dim} \mathfrak{h}$). 
Consider its $(N^2-1)$-dimensional adjoint representation labeled by a root system 
$\Delta=\Delta_+ \cup \Delta_-$, i.e. a set of vectors in a $r$-dimensional lattice. 
Given one positive (non-zero) root vector $\alpha\in\Delta_+$, 
we can choose $E_\alpha = E_{i,j} \in \mathfrak{n}_+$ ($i<j$). Note that 
$E_{i,j}$ is an $N\times N$ matrix with its $(i, j)$-th entry unity and zero otherwise. 
Take for example $\Phi={\rm diag}( x_1,\cdots,x_N )\in \mathfrak{h} ~({\rm subject~to}~x_1+\cdots+x_N=0$).%
\footnote{This constraint will correspond to decoupling the center of motion associated with 
the non-stationary $N$-body Lam\'{e} equation.} 
There holds 
\begin{eqnarray}
\label{30}
[  \Phi, E_\alpha ]= \alpha(\Phi) E_\alpha, ~~~~~~~~\alpha(\Phi)  =x_i -x_j .
\end{eqnarray}
Because all roots are located in a $r$-dimensional lattice and 
only $r$ of them are independent, let ${\bar \alpha}_a$ be simple roots such that 
$\alpha =\sum^r_{a=1} m_a {\bar \alpha}_a \in \Delta_+$ if 
$m_a\in {\bf Z}_{\ge 0}$. 
Generators in Cartan subalgebra are normalized 
 by the length constraint $(\alpha,\alpha)=2$ 
 where the inner product  is the usual one. 
In other words, given an orthonormal basis 
$\{ e_1,\cdots,e_N \}$ obeying $(e_{i}, e_{j})=\delta_{ij}$ let 
$\alpha \equiv e_{i}-e_{j} \in \Delta$ $(1\le i\ne j\le N)$. For positive roots 
in $\Delta_+$, $i<j$. For simple roots, $j=i+1$. Consequently, we are led to 
\begin{eqnarray*}
[ E_\alpha,  E_{-\alpha} ] =E_{i,i} -E_{j,j} \equiv H_\alpha, ~~~~~~~~ 
[ H_\alpha,  E_{\alpha} ]=\alpha (H_\alpha )E_\alpha=2E_\alpha,
\end{eqnarray*} 
which implies that 
the weight of each root vector is just encoded in $\Delta$.

From now on, we adopt the so-called Weyl-Cartan basis for $\mathfrak{g}$. That is, 
define ${\bf H}=(H_1, \cdots, H_{r}) \in \mathfrak{h}$ such that 
the highest weight state $|\lambda\rangle$ satisfies $H_a|\lambda\rangle=\lambda_a|\lambda\rangle$. 
When it comes to roots, for $\alpha = ( {\bf \alpha}_1, \cdots, {\bf \alpha}_r
)\in \Delta_+$ one has under this basis 
\begin{eqnarray}
\label{comm}
[ H_a, H_{b} ]=0, ~~~~~~~~ 
[ E_\alpha,  E_{-\alpha} ]= \sum^r_{a=1}
{\bf \alpha}_a^\vee  { H_a}, ~~~~~~~~ 
[ H_a ,  E_{\alpha} ]={\bf \alpha}_a E_\alpha.
\end{eqnarray}
Here, ${\alpha}^{\vee}=2\alpha /  (\alpha, \alpha)$ and 
 again $ (\alpha, \alpha)=2$ 
is imposed as a normalization of ${\bf H}$ where the inner
product $(\alpha, \alpha)=\sum_a {\alpha}_a \alpha_a$.

Let ${\bar \alpha}^\vee$'s be simple coroots represented by  
\begin{eqnarray*}
{\bar \alpha}^\vee_a =\sum^{r}_{b=1} A_{ab} \Lambda_b , ~~~~~~~~ 
 A_{ab}=( {\bar \alpha}_a, {\bar \alpha}_b^\vee): ~{\rm{ Cartan~matrix}}.
\end{eqnarray*} 
A set of fundamental weights $\{ \Lambda_a \}$ is used to express 
the highest weight vector as $\lambda=\sum_a \lambda_a \Lambda_a$ with 
$({\bar \alpha}_a^{\vee}, \Lambda_b)=\delta_{ab}$. 
The level $k$ of $\widehat{\mathfrak{sl}}_N$ is given by 
\begin{equation} 
k = \lambda_0 + (\theta, \lambda)=\sum^r_{a=0}  \lambda_a, ~~~~~~~~
\theta =\sum^r_{a=1} {\bf a}^{\vee}_a {\bar \alpha}_a^{\vee}, 
 =\sum^r_{a=1} {\bar \alpha}_a^{\vee},
~~~~~~~~{\bf a}^{\vee}_a:~{\rm{colabel}},\nonumber 
\end{equation}
where $\kappa=k+h^\vee$ ($h^\vee=\sum^r_{a=0} {\bf a}^{\vee}_a$: dual Coxeter number). 
For the $A$-type Lie algebra, ${\bf a}^{\vee}_{0}={\bf a}^{\vee}_a=1$. 
While the level $k$ goes to infinity, the integrable irreducible 
$\widehat{\mathfrak{g}}$-module reduces to that 
 of ${\mathfrak{g}}$.

\subsection{KZB equation}

We move to discuss correlation functions in WZW model. 
Of interest are their chiral parts, conformal blocks, 
satisfying Knizhnik-Zamolodchikov equations \cite{KZ,B}. 
 However, it is necessary for us to 
first get familiar with constructing Virasoro algebra from affine Lie ones.

Let $J^I_n$ ($n \in \mathbf{Z}$, $I,J=1,\cdots,{\rm dim}{\mathfrak{g}}$) be 
generators of 
the affine Lie algebra 
$\widehat{\mathfrak{g}}=\mathfrak{g} \otimes {\mathbf C}[z, z^{-1}] \oplus c{\mathbf C}$ 
whose central extension is $c={k {\text {dim}}\mathfrak{g}}/\kappa$
where 
\begin{eqnarray*}
12\rho^2 =h^\vee {\rm dim}{\mathfrak g}, ~~~~~~~~
\rho =\sum^r_{a=1} \Lambda_a=
\sum_{\alpha 
\in \Delta_+} \frac{\alpha}{2}:~{\rm Weyl~vector}. 
\end{eqnarray*}
According to Sugawara's construction, the generator of 
Virasoro algebra can be expressed via $J^I_n$:
\begin{eqnarray}
2\kappa 
L_n=
 \sum_{m\in {\mathbf Z}}g_{IJ}
:J_{n-m}^{I} J_{m}^{J}:
=g_{IJ} (
\sum_{m<0} J_{m}^{I} J_{n-m}^{J}
+\sum_{m \ge 0} J_{n-m}^{I} J_{m}^{J}
)\nonumber
\end{eqnarray}
where $g_{IJ}$ is the inverse of 
$g^{IJ}=K({\mathfrak{g}}_I ,{\mathfrak{g}}_J)$ called 
Killing form: 
\begin{eqnarray*}
K(H_a, H_b)=\delta_{ab}, ~~~~~~~~
K(E_\alpha, E_{-\alpha})=\frac{2}{\alpha^2} 
, ~~~~~~~~K(\cdot, \cdot)={\rm zero~otherwise}.
\end{eqnarray*}
In particular, $\kappa L_0$ is translated into 
 the quadratic Casimir operator 
$\Omega$ of ${\mathfrak{g}}$: 
\begin{eqnarray}
\label{el}
\Omega 
= \frac{1}{2} 
\sum_{a=1}^{r} 
H_a^2  +   \frac{1}{4} \sum_{\alpha\in \Delta_+}
{\alpha^2}
(E_\alpha E_{-\alpha}+ E_{-\alpha}E_{\alpha}), ~~~~~
\alpha^2=(\alpha, \alpha)
\end{eqnarray}
when applied to the highest weight state $|\lambda\rangle$. 
Following \eqref{comm}, we see that $L_0$ actually 
measures the conformal dimension 
$\Delta_\lambda$ of $|\lambda\rangle$ in the module $L_{k, \lambda}$: 
\begin{equation}
\label{conformal dimension}
\Delta_\lambda=\frac{(\lambda, \lambda+2\rho)}{2\kappa}
\end{equation}
where the inner product is taken w.r.t. the preceding Weyl-Cartan basis.

Our main concern are KZB equations which look like 
\begin{equation}
\label{kzbwzw}
\kappa \frac{\partial}{2\pi i\partial \tau'} \Psi =\frac{1}{4} H_0 \Psi , ~~~~~~
H_0 =-\frac{1}{2 \pi^2} \sum_{a=1}^{r} 
  \partial_{u_a} \partial_{u_a} 
  + 2 \sum_{\alpha \in \Delta} p(e^{\alpha(U)}) E_{\alpha} E_{-\alpha},
\end{equation}
where 
\begin{equation}
U=2\pi i ~\sum_{a=1}^{r}  u_a H_a, ~~~~~~
p(t)
 = - \sum_{m \in {\bf Z}} \frac{q'^{m} t}{(1-q'^{m} t)^2},~~~~~~q'=e^{2\pi i \tau'}
\end{equation}
and ${\alpha(U)}$ is defined in \eqref{30}. 
Certainly, \eqref{kzbwzw} is derived by applying the 
Ward-Takahashi identity associated with the 
energy-momentum tensor $T(z)=\sum_{n\in{\mathbf Z}} z^{-n-2} L_n$ to one-point 
toric conformal blocks $\Psi$: 
\begin{equation}
\label{cd1}
\Psi={\rm Tr}_{L_{k, \lambda}}~\Big(q'^{L_0-\frac{c}{24}} e^{U} v_{\ell}(z)\Big),  ~~~~~~
\ell \in P^k_+,  ~~~~~~
c=\frac{k {\text {dim}}\mathfrak{g}}{\kappa}.
\end{equation}
Notice that for $\mathfrak{sl}_2$ the primary field $v_{\ell} (z)$ is $V$-valued (taking its value in $V$) 
 and acted on by $\rho_{\ell}$ given certain spin-$\ell/2$ $\mathfrak{sl}_2$ 
 representation ($\rho_{\ell}, V_{\ell}$).
Still, the marked point $z$ located on the torus (complex moduli $\tau'$) can 
be sent to zero because $\Psi$ satisfies 
\be
\Big( z\frac{\partial}{\partial z}+\rho_{\ell}(\kappa^{-1}\Omega ) \Big) \Psi=0.   
\ee 
By definition, without any $v_{\ell} (z)$ 
inserted $\Psi$ reduces to 
the affine character associated with the integrable module 
$L_{k, \lambda}$:
\begin{equation}
\Psi \to \chi={\rm Tr}_{L_{k, \lambda}}~\Big(q'^{L_0-\frac{c}{24}} e^{U} \Big).
\end{equation}

Let us pause for a while 
to discuss the $V$-valuedness of $v_{\ell} (z)$. 
This can be done twofold. 
First, define $v(z)\equiv v(\zeta|z)$ which depends additionally on an internal coordinate 
$\zeta$. 
Its OPE with some $\widehat{\mathfrak{sl}}_2$-current field $J^I (w)$ reads
\begin{equation*}
J^I (w)v_{}(\zeta|z) \sim \frac{1}{w-z} {\cal D}^I(\zeta)v_{}(\zeta|z), 
 ~~~~~~{\cal D}^I (\zeta)v_{}(\zeta|z)\equiv 
\rho(J^I)v(z).
\end{equation*}
Second, we resort to Wakimoto's representation. 
Introduce the  primary field corresponding to the $\widehat{\mathfrak{sl}}_2$
highest weight state $| \ell \rangle$ in terms of a chiral free boson $\varphi (z)$: 
\be
\label{defi}
| \ell \rangle ~\to~ :\exp (\ell /\sqrt{2}) \varphi (z): \nonumber
\ee 
whose conformal dimension $\Delta_\ell$ is just 
computed in \eqref{conformal dimension} with $\lambda\to\ell$. 
Instead of the additional $\zeta$-dependence, one 
prepares another chiral free field $\gamma(z)$%
\footnote{Note that ${\bf dim}(\Delta_+)$ is equal to the 
 total number of pairs of $\big( \beta(z), \gamma(z)\big)$.} 
 and constructs the full ${\mathfrak{sl}}_2$ 
 spin-$\ell/2$ multiplet which contains 
\begin{equation}
\label{other}
 \gamma(z)^{\ell/2-m} :\exp (\ell /\sqrt{2}) \varphi (z):,  
 ~~~~~~m=-\ell/2,\cdots, \ell/2.
\end{equation}
Then, $v_{\ell}(z)$ can be identified with one of them. 

From \eqref{cd1} we realize that the role of 
$v_{\ell} (z)$ is an intertwiner, i.e. $v_{\ell} (z):~L_{k, \lambda}\to  L_{k, \lambda}\otimes V[0]$. 
Here, $V[0]$ stands for the one-dimensional 
zero-weight subspace of $V_{\ell}$, ($\ell+1$)-dimensional  ${\mathfrak{sl}}_2$-module.%
\footnote{$V[0]$ vanishes if ${\bf dim}(V)=\ell+1$ $(\ell=0,1,2,\cdots)$ is even.} 
Due to the 
Ward-Takahashi identity w.r.t. $\widehat{\mathfrak{sl}}_2$-current fields applied to $\Psi$, we see 
($H$: ${\mathfrak{sl}}_2$ Cartan generator)
\begin{equation}
\label{even}
\rho_{\ell}(H)\Psi=0.
\end{equation}
This explains why $v_{\ell}(z)$ belongs to the zero-weight subspace $V[0] \subset V_{\ell}$.

\subsection{$H_0$}

Let us describe $H_0$ in \eqref{kzbwzw} 
in more detail. We want to look into the function $p(t)$ 
\cite{KT1,KT2,96, FSV} inside $H_0$. 
By using Weierstrass's $\wp$-function it gets expressed by ($t = e^{2\pi i w}$)
\begin{align}
\label{wpwp}
&p(t) = -\frac{t}{(1-t)^2} 
 - \sum_{m\ne 0} 
\frac{q'^{m} t}{(1 - q'^{m}t)^2} , \nonumber\\
&4\pi^2 p(t) = -\partial_w^2\log \theta_1 (w|q'^{\frac{1}{2}}) 
=\wp(w|q'^{\frac{1}{2}}) + 2\eta_1(q'^{\frac{1}{2}}).
\end{align}
Here, $\wp$, $\theta_1$ and $\eta_1$ follow the same 
convention adopted in section \ref{sec2}. 
Eq. \eqref{wpwp} is explained as below. 
Because $p(t)$ has order-two poles 
at $ \{q'^{m}\}_{m\in{\bf Z}}$ and satisfies the periodicity condition 
$p(q't) = p(t)$, it can be rewritten into the form 
\be
 p(t) = -\frac{t}{(1-t)^2} 
 - \sum_{m>0} \left\{ 
 \frac{q'^{m} t}{(1 - q'^{m}t)^2} + \frac{q'^{m} t^{-1}}{(1 - q'^{m} t^{-1})^2} 
 \right\}. \nonumber
\ee
Recalling $t = e^{2 \pi i w}$ we find the first term becomes
\be
 - \frac{t}{(1 - t)^2} = \frac{1}{4\sin^2(\pi w)},  \nonumber
\ee
and other terms become
\be
 - \frac{q'^{m} t}{(1 - q'^{m}t)^2} - \frac{q'^{m} t^{-1}}{(1 - q'^{m} t^{-1})^2}
= - 2 q'^m \frac{\cos(2 \pi w) (1 + q'^{2m}) - 2 q'^m}
{\big(1 - 2 q'^m \cos (2 \pi w) + q'^{2m}\big)^2}. \nonumber
\ee
On the other hand, based on the product representation of $\theta_1(w|q'^{\frac{1}{2}})$: 
\be
 \theta_1(w|q'^{\frac{1}{2}}) = 2 q'^{\frac{1}{8}} \sin(\pi w) 
 \prod_{m>0} (1 - q'^m) \big(1 - 2 q'^m \cos (2 \pi w)   + q'^{2m}\big)\nonumber
\ee
we have
\be
 - {\p_w^2} \log \theta_{1}(w|q'^{\frac{1}{2}}) 
 = \frac{\pi^2}{\sin^2(\pi w)} 
 - \sum_{m>0} 8 \pi^2 q'^m \frac{\cos(2 \pi w) (1 + q'^{2m}) 
 - 2 q'^m}{\big(1- 2 q'^m \cos (2 \pi w) + q'^{2m} \big)^2}.  \nonumber
\ee
Combined with 
\be
 - {\p_w^2} \log \theta_{1}(w|q'^{\frac{1}{2}})  
 = \wp(w|q'^{\frac{1}{2}}) + 2 \eta_1(q'^{\frac{1}{2}}), \nonumber
\ee
we go back to \eqref{wpwp}. 
To explicitly evaluate $e^{\alpha(U)}$ for $p(t)$, 
we resort to \eqref{30}. Generally, in the case of $\mathfrak{sl}_N$ 
\begin{equation*}
\frac{1}{2\pi i}U=\sum^r_{a=1} u_a H_a={\rm diag}(y_1,\cdots,y_N),  ~~~~~~\sum^N_{i=1} y_i=0.
\end{equation*}
Due to $(\alpha, \alpha)=2$ as stressed, one sees 
$\alpha(U)=2\pi i \sqrt{2} u_1$ and $w\to \sqrt{2} u_1\equiv u$ 
for $\mathfrak{sl}_2$.

Finally, we want to determine the eigenvalue of $E_\alpha E_{-\alpha }$ in 
\eqref{kzbwzw} which 
acts on $V[0]\subset V_{\ell}$ 
of the primary field $v_{\ell}(z)$. 
Since $V[0]={\bf C}$ is one-dimensional, in view of \eqref{other} 
one can assume that it is spanned by some monomial like $(L_1 \cdots L_{N})^{\xi}$. 
Furthermore, 
for $\mathfrak{sl}_N$ 
there exists the following representation: 
\begin{equation*}
E_\alpha \equiv E_{ij} = 
L_i \frac{\partial}{\partial L_j},  ~~~~~~
i\ne j=1,\cdots N,  ~~~~~~\alpha\in \Delta
\end{equation*}
whereas 
\begin{equation*}
H_i=L_i \frac{\partial}{\partial L_i}
- L_{i+1} \frac{\partial}{\partial L_{i+1}},  ~~~~~~
i=1,\cdots N-1.
\end{equation*}
In the case of $N=2$ (or $\mathfrak{sl}_2$) 
 we thus obtain the eigenvalue $ \ell(\ell+2)/4$ of $E_\alpha E_{-\alpha }$
 through $\xi  \equiv \ell/2$. 
This choice of $\xi$ is rigid and not arbitrary.

\subsection{Comparison}
Equipped with these, we are in a position to replace 
$H_0$ in \eqref{kzbwzw} by  
\begin{equation*}
\label{detail}
 H_{0} = \frac{1}{\pi^2}\left\{-\frac{\p^2}{\p u^2} 
 + \frac{\ell(\ell+2)}{4}  \Big( \wp(u|q'^{\frac{1}{2}}) +2 \eta_1(q'^{\frac{1}{2}}) \Big) \right\}, 
 ~~~~~~u\equiv \sqrt{2}u_1.
\end{equation*}
We then arrive at the familiar form of KZB equations: 
\begin{align}
\kappa \frac{\partial}{2\pi i\partial \tau'} \Psi
=   \frac{1}{4\pi^2} \left\{-\frac{\p^2}{\p u^2} 
 + \frac{\ell(\ell+2)}{4}  \Big( \wp(u|q'^{\frac{1}{2}}) 
 + 2 \eta_1(q'^{\frac{1}{2}}) \Big) \right\} \Psi.
\label{kzbheat}
\end{align}
Let us slightly rewrite \eqref{kzbheat} into 
\begin{align}
2 \pi \kappa 
\frac{\partial}{ i\partial \tau'} \widetilde{\Psi}
=   \left\{-\frac{\p^2}{\p u^2} 
 + \frac{\ell(\ell+2)}{4}   \wp(u|q'^{\frac{1}{2}})  \right\} \widetilde{\Psi}
\label{kzbheat1}
\end{align}
with 
\begin{align}
\widetilde{\Psi}=\exp\Big(-\frac{i}{2\pi \kappa}\frac{\ell(\ell+2)}{4}\int^{\tau'}   2\eta_1 d\tau'' \Big)\Psi .
\label{henkei}
\end{align} 
Remark that by $(\tau', u) \to \kappa^{-1}(\tau', u)$ and $\widetilde{\Psi}\to \Phi$ (see footnote 3) 
\eqref{kzbheat1} becomes 
\begin{align}
2 \pi
\frac{\partial}{ i\partial \tau'} \Phi
=   \left\{-\frac{\p^2}{\p u^2} 
 + \frac{\ell(\ell+2)}{4}   \wp(u|q'^{\frac{1}{2}})  \right\} \Phi.
\label{kzbheat2}
\end{align}

Exactly the same procedure of redefining the wave 
function 
$\Psi_{\pm}^{(8 \mathrm{vertex})}$ by absorbing $\eta_1$ into it can be applied to \eqref{TT-Lame2}. After doing that, comparing 
\eqref{TT-Lame2} with 
\eqref{kzbheat} we find 
\begin{align}
\label{first}
s \Longleftrightarrow u, ~~~~~~
n \Longleftrightarrow \ell/2, ~~~~~~
\bar{\tau} \Longleftrightarrow \tau^\prime. 
\end{align}
The relationship \eqref{first} reveals 
 that keeping the total spin-chain site number $M$ odd \cite{TT-St} 
can now be interpreted as the existence requirement 
 for one-point toric conformal blocks due to $M=2n+1={\bf dim}V$ 
in view of \eqref{even}. This serves as a geometric interpretation of Stroganov's claim. 
We will provide another consistency check in section \ref{sec4}.

\section{Test: reduction to Sutherland model}\label{sec4}

\begin{figure}[tt]
\begin{center}
\includegraphics[width=0.8\textwidth]
{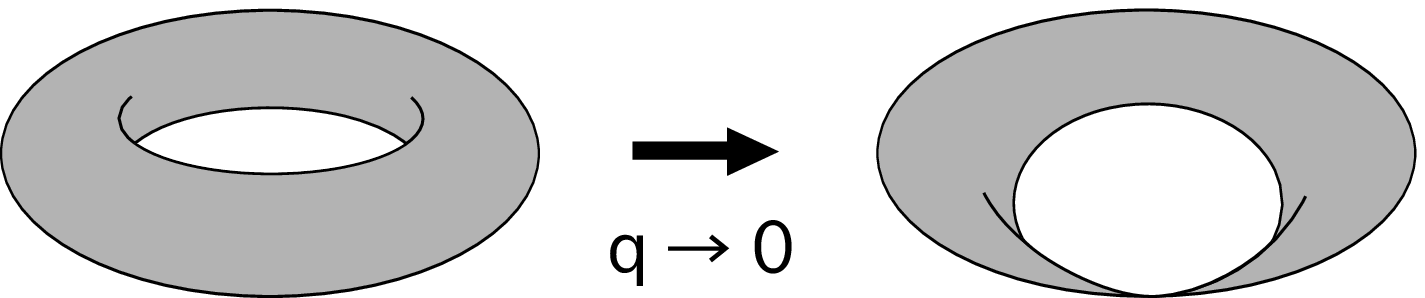}
\end{center}
\caption{As $q\to0$ one has a degenerate torus. 
\label{F3}}
\end{figure}

Performing a trigonometric 
reduction to the spin-1/2 XXZ chain 
($q \to 0$) helps 
strengthen the correspondence indicated in \eqref{first}. 
On WZW model side, this leads to 
a degenerate torus drawn in Fig. \ref{F3}.%
\footnote{See also \cite{Tai:2010im, Tai:2010ps} where the issue 
presented here was encountered within the context of 2d 
Liouville CFT/4d ${\cal N}=2$ gauge theory 
correspondence initiated in 
\cite{Nekrasov:2009ui, Alday:2009aq, Nekrasov:2009rc, Nekrasov:2011bc}.}

Based on $c(0,n)=3n(n+1)$ and
\begin{align}
\Psi_{\pm}^{(8 \mathrm{vertex})}(s/3,q,n)=q^{\frac{3}{2}(d_{\pm}+\frac{1}{4})}
 \Psi_{\pm}^{(6 \mathrm{vertex})}(s/3,n)\big( 1+{\cal O}(q)\big), ~~~~~~
d_{\pm}=\frac{1 \mp 6}{36}, \nonumber
\end{align}
one finds that eq. 
\eqref{TT-Lame2} becomes 
\[
\left\{-\frac{\p^2}{\p s^2} 
 + n(n+1) \left( \wp(s|0) +\frac{1}{3} \right)
- \pi^2 \left( d_{\pm}+ \frac{1}{4} \right)
 \right\} \Psi_{\pm}^{(6 \mathrm{vertex})}(s/3,n)=0.
\]
Furthermore, from
\begin{align}
\wp(s|0)=\frac{\pi^2}{\mathrm{sin}^2 \pi s}-\frac{1}{3},~~~~~~
\frac{\theta_1^\prime(s|q)}{\theta_1(s|q)}~
 \xrightarrow[q \to 0]{}~ \pi \frac{\mathrm{cos} \pi s}{\mathrm{sin} \pi s },
\nonumber
\end{align}
we arrive at the two-body Sutherland model:
\be
\left\{-\frac{\p^2}{\p s^2} 
 + \frac{\pi^2 n(n+1)}{\mathrm{sin}^2 \pi s}
-\pi^2 \left( d_{\pm}+ \frac{1}{4} \right)
 \right\} \Psi_{\pm}^{(6 \mathrm{vertex})}(s/3,n)=0.
\label{Sutherland}
\ee
Then, through%
\footnote{Another transformation: 
\begin{align}
\widetilde{\Psi}_{\pm}^{(6 \mathrm{vertex})}(s)=
(\mathrm{sin} \pi s)^n 
\Psi_{\pm}^{(6 \mathrm{vertex})}(s/3,n)\nonumber
\end{align}
will lead to Stroganov's result \cite{TT-St}.  
} 
\begin{align}
\widetilde{\Psi}_{\pm}^{(6 \mathrm{vertex})}(s)=
(\mathrm{sin} \pi s)^{-n-1}
\Psi_{\pm}^{(6 \mathrm{vertex})}(s/3,n)\nonumber
\end{align}
we are able to rewrite \eqref{Sutherland} into ($\pi s={\tilde s}$)
\begin{align*}
\left\{-
\frac{\p^2}{\p {\tilde s}^2} 
-2(n+1) \mathrm{cot} {\tilde s} \frac{\p}{\p {\tilde s}}
 + (n+1)^2 - \left(d_{\pm}+ \frac{1}{4} \right)
 \right\} \widetilde{\Psi}_{\pm}^{(6 \mathrm{vertex})}(s)=0.
\end{align*}
In fact, 
$\widetilde{\Psi}_{\pm}^{(6 \mathrm{vertex})}(s)$ is related to the 
Gegenbauer polynomial $G^{(\nu)}_{e}(\cos {\tilde s})$: 
\begin{align}
G^{(\nu)}_{e}(\cos{\tilde s})=
\frac{\Gamma(e+2\nu)}{\Gamma(2\nu) e!}~ 
{}_2 F_1 \left(-e,e+2\nu,\nu+\frac{1}{2};\frac{1-\cos{\tilde s}}{2}\right)
\nonumber
\end{align}
through $\nu=n+1$ and $e(e+2\nu)=(d_{\pm}+ \frac{1}{4})-(n+1)^2$.

On the other hand, it is also known that 
$\widetilde{\Psi}_{\pm}^{(6 \mathrm{vertex})}(s)$ 
by changing the variable to 
$S=\exp ( i {\tilde s})$ becomes 
the Jack polynomial $J^{(\nu)}_{\lambda}(S)$ 
 where $\lambda$ denotes the energy level. 
More precisely, by $\Delta = \left(\sin \tilde{s} \right)^{n+1}$
the Hamiltonian $H_S$ of the two-body Sutherland model 
 is transformed into 
\[
 H_0 = \Delta^{-1} (H_S - e_0) \Delta, ~~~~~~
 H_{S} =  -\frac{\p^2}{\p \tilde{s}^2} + 
 \frac{n(n+1)}{\sin^2 \tilde{s}},  
\]
whose eigenfunction is the Jack polynomial $J_{\lambda}^{(\nu)}(S)$. 
Here, $e_0$ stands for the eigenvalue of $H_0$ w.r.t. the ground-state $\Delta$. 
In addition, 
as $n \to 0$ 
$J^{(\nu)}_{\lambda}(S)$ 
reduces to the Schur function: 
\begin{align}
\chi_{\lambda} (S)=\frac{S^{1+\lambda} - S^{-1-\lambda}}{S - S^{-1} }.
\label{Schur}
\end{align}
Eq. \eqref{Schur} has its ${\mathfrak{sl}}_N$ analogy, i.e. 
given ${\bf S}={\rm diag}(S_1, \cdots, S_N)$ ($\det {\bf S}=1$) one has 
\begin{align*}
\chi_{R}({\bf S})=\frac{\det (S_i^{R_j+N-j})}{\det(S_i^{N-j})}, ~~~~~~i,j=1,\cdots,N,
\end{align*}
where $R=(R_1,\cdots,R_{N-1},0)$ stands for a Young tableau. 
Each row length of $R$ obeys $R_1 \ge R_2 \ge \cdots$. 
In addition, there exists 
\begin{align*}
( R_1-R_2,\cdots, R_i - R_{i+1},\cdots, R_{N-1} )=
(\lambda_1, \cdots, \lambda_i,\cdots,\lambda_{N-1})
\end{align*}
between $R$ and $\lambda \in P_+$ of ${\mathfrak{sl}}_N$. 

To see the emergence of \eqref{Schur} on CFT side, 
we follow two steps below whose order differs from the above procedure.\\
${\bf Step~1}$: As mentioned in section \ref{sec3}, when the insertion becomes an 
identity operator 
($\ell \to 0$) 
the toric conformal block reduces to the level-$k$ $\widehat{\mathfrak{sl}}_2$ character which 
is explicitly ($k\ge \lambda$, $\lambda$: highest weight)
\begin{align*}
\chi&=\frac{\theta_{\sqrt{2}\lambda+1,k+2}-\theta_{-\sqrt{2}\lambda-1,k+2}}
{\theta_{1,2}-\theta_{-1,2}}, \\
\theta_{A,B}&\equiv \theta_{A,B} (u_1|q')=\sum_{n \in \mathbf{Z}+\frac{A}{2B}}
(q^\prime)^{ B n^2}
\mathrm{e}^{2 B n\pi i  u_1}.
\end{align*}
\\
${\bf Step~2}$: Next, we take $\tau^\prime \to i\infty$ such that only 
those terms involving ${\bf Z}=0$ inside the summation of $\theta_{A,B} $ survive. 
The above affine character $\chi$ factorizes
into two parts depending on respectively $q^\prime$ and $u_1$:
\begin{align*}
\chi ~ \xrightarrow[q^\prime \to 0]{} ~{q'}^{\big(  \frac{( \sqrt{2}\lambda+1)^2}{4 \kappa} -\frac{1}{8} \big)}
\frac{\mathrm{e}^{\pi  i (\sqrt{2}\lambda+1)u_1}-\mathrm{e}^{-\pi  i (\sqrt{2}\lambda+1)u_1}}
{\mathrm{e}^{\pi  i u_1}-\mathrm{e}^{-\pi  i u_1}}.
\end{align*}
Actually, the factor $\frac{( \sqrt{2}\lambda+1)^2}{4\kappa} -\frac{1}{8}$ 
comes from $L_0-\frac{c}{24}$. Because 
when the degeneration depicted in Fig. \ref{F3} occurs only the highest weight state 
in the integrable module $L_{k, \lambda}$ contributes to $\chi$. 
After dropping the 
$q^\prime$-dependent part we are simply left with ($\rho=1/\sqrt{2}$: Weyl vector)  
\begin{align*}
\Psi ~ \xrightarrow[\ell \to 0]{} ~ \chi  ~\xrightarrow[q^\prime \to 0]{}~
\frac{\mathrm{e}^{\pi  i (\sqrt{2}\lambda+1)u_1}-\mathrm{e}^{-\pi  i (\sqrt{2}\lambda+1)u_1}}
{\mathrm{e}^{\pi  i u_1}-\mathrm{e}^{-\pi  i u_1}}
=\frac{\mathrm{e}^{\pi  i (\lambda+\rho)u}-\mathrm{e}^{- \pi  i (\lambda+\rho)u }}
{\mathrm{e}^{\pi  i \rho u}-\mathrm{e}^{-\pi  i \rho u}}.
\end{align*}
This process we are studying then serves as a confirmation of the link 
 (Table \ref{dic}) 
between WZW model and the spin-1/2 XYZ chain.

\begin{figure}[tt]
\begin{center}
\includegraphics[width=0.8\textwidth]
{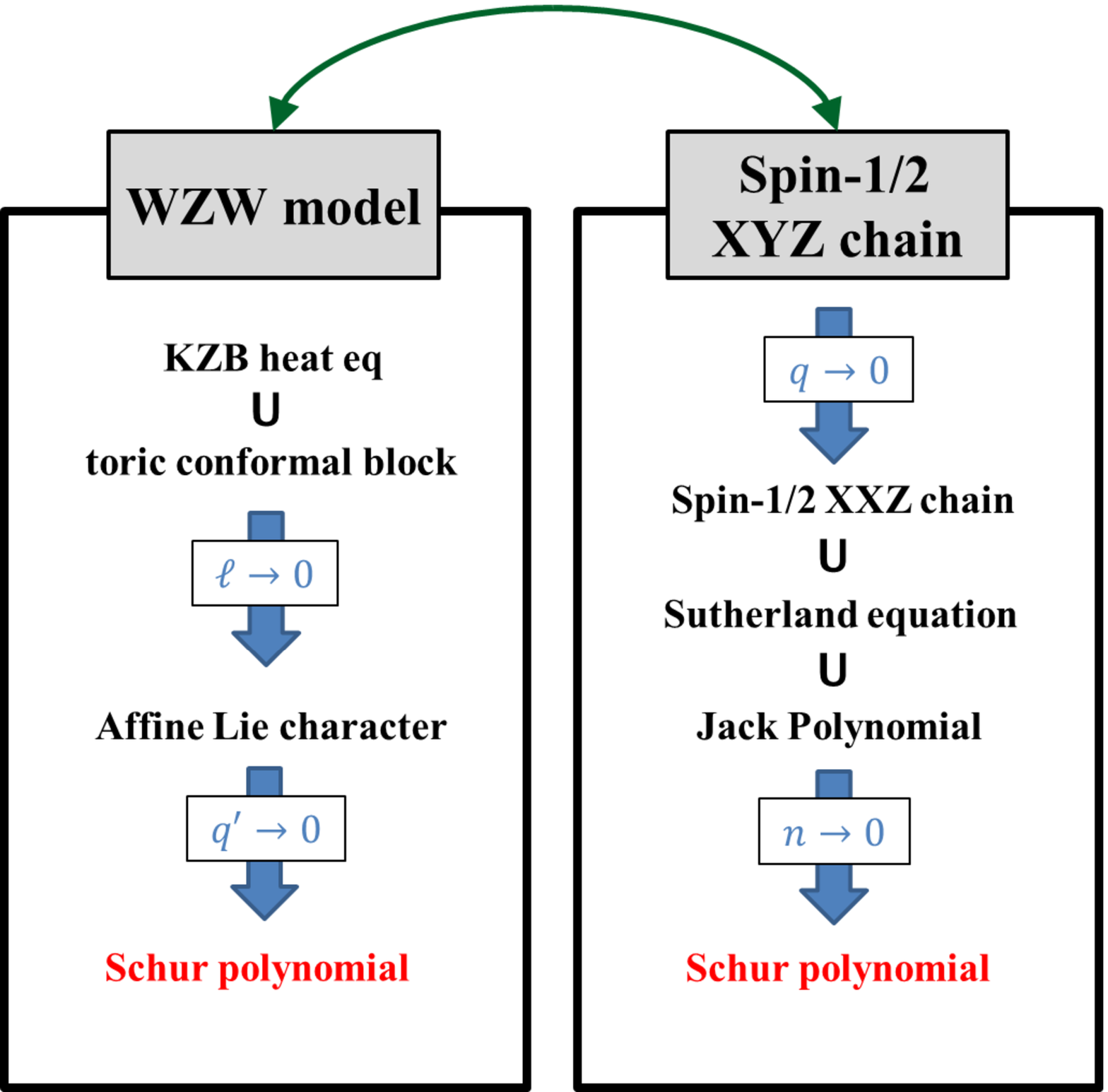}
\end{center}
\caption{${\bf (RHS)}$ A trigonometric degeneration of 
the spin-1/2 XYZ chain. ${\bf (LHS)}$ Through the identification 
 $\ell/2=n$, toric conformal blocks 
reduce correspondingly to Schur polynomials 
 (eigenstates of the two-body Sutherland model in RHS as $n\to 0$). 
Note that the order of two limits differs between RHS and LHS.
\label{F2}}
\end{figure}

\section{Summary} 
We have provided an interpretation of Stroganov's ``The Importance of being Odd" 
at the Razumov-Stroganov point by means of CFT language. 
We found that the total number of the XYZ chain sites 
$M=2n+1$ is equal to the dimension of 
the ${\mathfrak{sl}}_2$ 
representation space $V$ w.r.t. the primary field inserted on a torus in WZW model. 
Notice that $M$ must be odd. 
The approach summarized in Fig. \ref{F2} was used to support our proposal. 

In fact, there is still another interesting limit mentioned in \cite{BM}, 
i.e. $n\to \infty$ and $q\to 0$ with $t=8 q^{\frac{3}{2}} n$ kept fixed as 
presented in Appendix. 
To study the corresponding deformation of WZW conformal blocks is 
an interesting future work, though the explicit form of them is not available. 
In the context of 2d Liouville field theory characterized by Virasora algebra, similar issues 
have been addressed in \cite{Mironov:2009uv, Alba:2009fp}.

\section*{Acknowledgments}
We thank professor Hratchya M. Babujian for helpful comments. 

\section*{Appendix}
Under $ n\to \infty$ and $q\to 0$ with $t=8 q^{3/2} n$ kept fixed, 
one has $\Psi^{(8 \text{vertex})}_{\pm} \to \mathcal{Q}_{\pm} (\theta,t)$.

By further adopting the new variable $\theta$ defined by 
 $ i\theta=s-\frac{1}{2}\pi \bar{\tau}$, 
this limit applied to \eqref{TT-Lame1} leads to a massive sine-Gordon model 
 on a cylinder 
\cite{BM}: 
\be
 t \frac{\p}{\p t} \mathcal{Q}_{\pm} (\theta, t) = 
 \left\{ \frac{\p^2}{\p \theta^2} - \frac{1}{8} t^2 (\cosh 2\theta - 1) 
 \right\} \mathcal{Q}_{\pm}(\theta,t). \nonumber
\ee

\end{document}